\documentclass[final,english]{bullsrsl}

\usepackage[utf8]{inputenc}
\usepackage[T1]{fontenc}
\usepackage{natbib}
\usepackage{graphicx}

\begin{document}
\title{Solving the mystery of extreme light variability in the massive eccentric system MACHO\,80.7443.1718}

\author[affil={1,2},corresponding]{Piotr A.}{Kołaczek-Szymański}
\author[affil={1}]{Piotr}{Łojko}
\author[affil={1}]{Andrzej}{Pigulski}
\author[affil={1,3}]{Tomasz}{Różański}
\author[affil={1}]{Dawid}{Moździerski}
\affiliation[1]{Uniwersytet Wrocławski, Wydział Fizyki i Astronomii, Instytut Astronomiczny, ul. Kopernika 11, 51-622 Wrocław, Poland}
\affiliation[2]{Space sciences, Technologies and Astrophysics Research (STAR) Institute, Universit\'e de Li\`ege, All\'ee du 6 Ao\^ut
19c, B\^at.~B5c, 4000 Li\`ege, Belgium}
\affiliation[3]{Australian National University, Research School of Astronomy\,\&\,Astrophysics, Cotter Rd., Weston, ACT 2611, Australia}
\correspondance{piotr.kolaczek-szymanski@uwr.edu.pl}
\date{16th October 2024}
\maketitle

\begin{abstract}
The evolution of massive stars is heavily influenced by their binarity, and the massive eccentric binary system MACHO\,80.7443.1718 (ExtEV) serves as a prime example. This study explores whether the light variability of ExtEV, observed near the periastron during its 32.8-day orbit, can be explained by a wind-wind collision (WWC) model and reviews other potential explanations. Using broadband photometry, TESS data, ground-based $UBV$ time-series photometry, and high-resolution spectroscopy, we analysed the system’s parameters. We ruled out the presence of a Keplerian disk and periodic Roche-lobe overflow. Our analysis suggests the primary component has a radius of about $30\,{\rm R}_\odot$, luminosity of $\sim6.6\times10^5\,{\rm L}_\odot$, and mass between $25$ and $45\,{\rm M}_\odot$, with a high wind mass-loss rate of $4.5\times10^{-5}\,{\rm M}_\odot\,{\rm yr}^{-1}$, likely enhanced by tidal interactions, rotation, and tidally excited oscillations. We successfully modelled ExtEV’s light curve, identifying atmospheric eclipse and light scattering in the WWC cone as key contributors. The system's mass-loss rate exceeds theoretical predictions, indicating that ExtEV is in a rare evolutionary phase, offering insights into enhanced mass loss in massive binary systems.
\end{abstract}

\keywords{close binary systems, early-type stars, emission-line stars, massive stars, mass-loss, MACHO\,80.7443.1718}

\section{Introduction}
\begin{figure}[ht]
\centering
\includegraphics[width=\hsize]{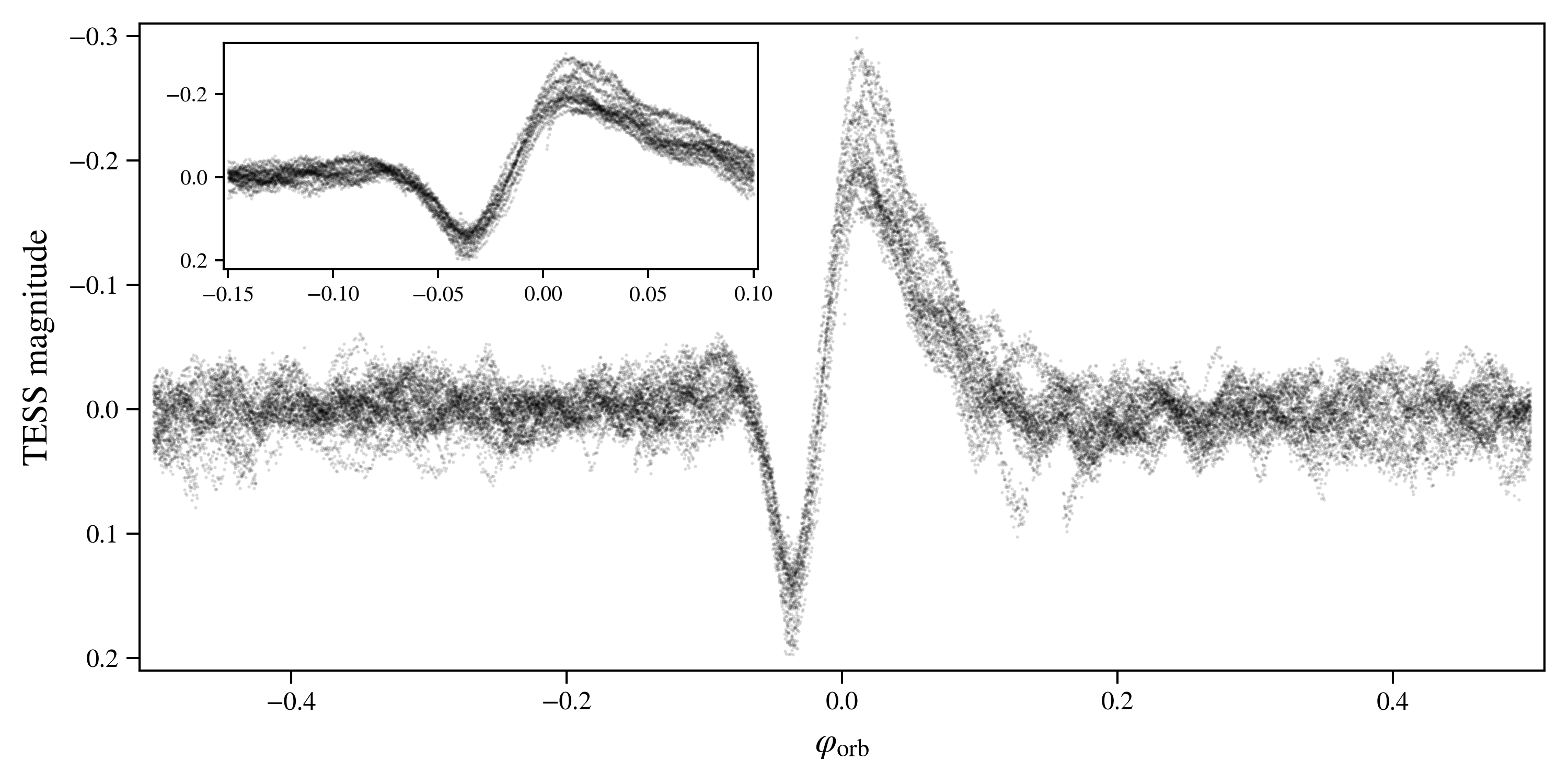}
\bigskip
\begin{minipage}{12cm}
\caption{TESS near-infrared light curve of ExtEV phased with the orbital period of 32.83016 d. Orbital phase $\varphi_{\rm orb}=0$ corresponds to the time of the periastron passage. The inset shows the light curve near the periastron passage. Stochastic light variations and TEOs can be seen over the entire range of orbital phases.}
\end{minipage}
\end{figure}
Most massive stars reside in binary and multiple systems \citep[e.g.,][]{1998NewA....3..443V,2012Sci...337..444S,2013ARA&A..51..269D}. The presence of a relatively close companion can affect the mass loss rate (MLR) due to line-driven stellar wind \citep[e.g.,][]{2020ApJ...902...85M} or induce high-amplitude tidally excited oscillations \citep[TEOs; e.g.,][]{2017MNRAS.472.1538F}. In particular, the intensity of stellar wind has fundamental implications for the evolutionary track of a massive star \citep[e.g.,][]{2014ARA&A..52..487S,2022ARA&A..60..203V,2023A&A...676A.109B}. Unfortunately, there are still many uncertainties regarding theoretical predictions of MLRs in massive stars due to their clumped, radiation-driven winds, especially when they leave the main sequence and become blue supergiants \citep[BSGs; e.g.,][]{2023A&A...673A.109G}. Therefore, any opportunity to observe massive binary systems during the ongoing process of envelope stripping due to mass transfer or stellar wind enhanced by the presence of a companion is particularly valuable for verifying theoretical studies. In this context, MACHO\,80.7443.1718 is of particular interest.

MACHO 80.7443.1718 (ExtEV, which stands for an ``extreme eccentric variable'') is an eccentric and massive binary (or even triple hierarchical) system with an orbital period of approximately 33\,d, located in the LMC \citep{2021MNRAS.506.4083J}. The primary component of this system is a BSG of spectral type B0\,Iae, while the secondary component is likely an O-type dwarf. ExtEV is a unique object for at least two reasons. Firstly, its light curve (without photospheric eclipses) exhibits an exceptionally large range of flux changes near periastron ($\sim 0.4$ mag; Fig.~1), which initially led \cite{2019MNRAS.489.4705J} to reclassify the object as an ``extreme heartbeat star''. However, it quickly became apparent that it could not be the case as the model of ellipsoidal variability fails to reproduce the photometric amplitude of ExtEV. The actual source of ExtEV's variability became intriguing enough for the system to be featured in a dedicated publication in Nature Astronomy \citep{2023NatAs...7.1218M}. The authors suggest that the system
experiences periodic Roche-lobe overflow (RLOF) near the periastron, accompanied by the nonlinear tidal wave breaking on the surface of the primary component \citep{2022ApJ...937...37M}. However, this model suffers from numerous assumptions and approximations, making it questionable \citep{2024A&A...686A.199K}. The second distinguishing feature of ExtEV is the presence of high-amplitude TEOs. It is also the first system where relatively sudden and strong changes in the amplitude of TEOs have been detected \citep{2022A&A...659A..47K}. Therefore, we revised the existing models of ExtEV variability and proposed another solution, based on the new spectroscopic and photometric data collected by us \citep{2024A&A...686A.199K}. This article is a short summary of the extensive analysis that we describe in detail in the aforementioned paper.

\section{Key properties of the ExtEV system and its primary component}
Using the Las Cumbres Observatory Global Telescope (LCOGT) light curves of ExtEV in the Johnson $U$, $B$, and $V$ passbands, we conclude that the peak-to-peak amplitude and the shape of the ExtEV's light curve seem to be independent of wavelength over a broad range from near-ultraviolet to near-infrared (IR). The spectral energy distribution (SED) of ExtEV reveals a significant IR excess, especially in the mid-IR bands (WISE W3 and W4 filters). Although the IR properties of ExtEV suggest an increased mass loss rate in the system, they do not align with the typical characteristics of B[e] SGs, as no emission lines like [O\,I] lines \citep[which are typical of the Be phenomenon;][]{1998A&A...340..117L} were detected in our SALT High-Resolution Spectrograph (SALT/HRS) spectra. Our SED modeling indicates that the primary component has a radius of approximately $30\,{\rm R}_\odot$ and a luminosity of about $\log(L/{\rm L}_\odot)\approx 5.82$ with an additional attenuation around 2760\,\AA\ of unknown origin. Determining the initial and current mass of the primary component is challenging due to the impact of rotation and mass loss on its evolution, but simulations we performed with the MESA software \citep{2011ApJS..192....3P,2023ApJS..265...15J} suggest the zero-age main-sequence mass likely falls between $27$ and $55\,{\rm M}_\odot$, with a present mass between $25$ and $45\,{\rm M}_\odot$. The evolutionary status of the primary component is difficult to define. It is either in the hydrogen or helium core-burning phase. We also derived spectroscopic parameters of the primary's orbit by combining archival radial-velocity data with those from SALT/HRS spectra, obtaining a more precise mass function of $f(M)=0.74\pm0.05\,{\rm M}_\odot$ and identifying significant changes in the systemic velocity of ExtEV, suggesting that it may be a hierarchical triple system. Our analysis shows that assuming the H$\alpha$ and H$\beta$ emission features arise from a hypothetical Keplerian disk around the primary component leads to a disk location that would strongly interact with the secondary’s orbit. Furthermore, the primary component likely maintains a detached geometry during periastron passage, with a rotation period of about 8 days, consistent with pseudo-synchronous rotation. The light curve of ExtEV can be explained by a superposition of an atmospheric eclipse of the secondary by the primary's intense stellar wind and excess emission from scattered light on the dense structures of the WWC cone. Using the theoretical model of atmospheric eclipses developed by \cite{1996AJ....112.2227L} and later modified by \cite{2021A&A...650A.147S}, we determined the orbital inclination of ExtEV to be approximately $66^\circ$ and the wind mass-loss rate from the primary component as large as $4.5\times10^{-5}\,{\rm M}_\odot\,{\rm yr}^{-1}$. The results of this modelling are depicted in Fig.~2.
\begin{figure}[ht]
\centering
\includegraphics[width=\hsize]{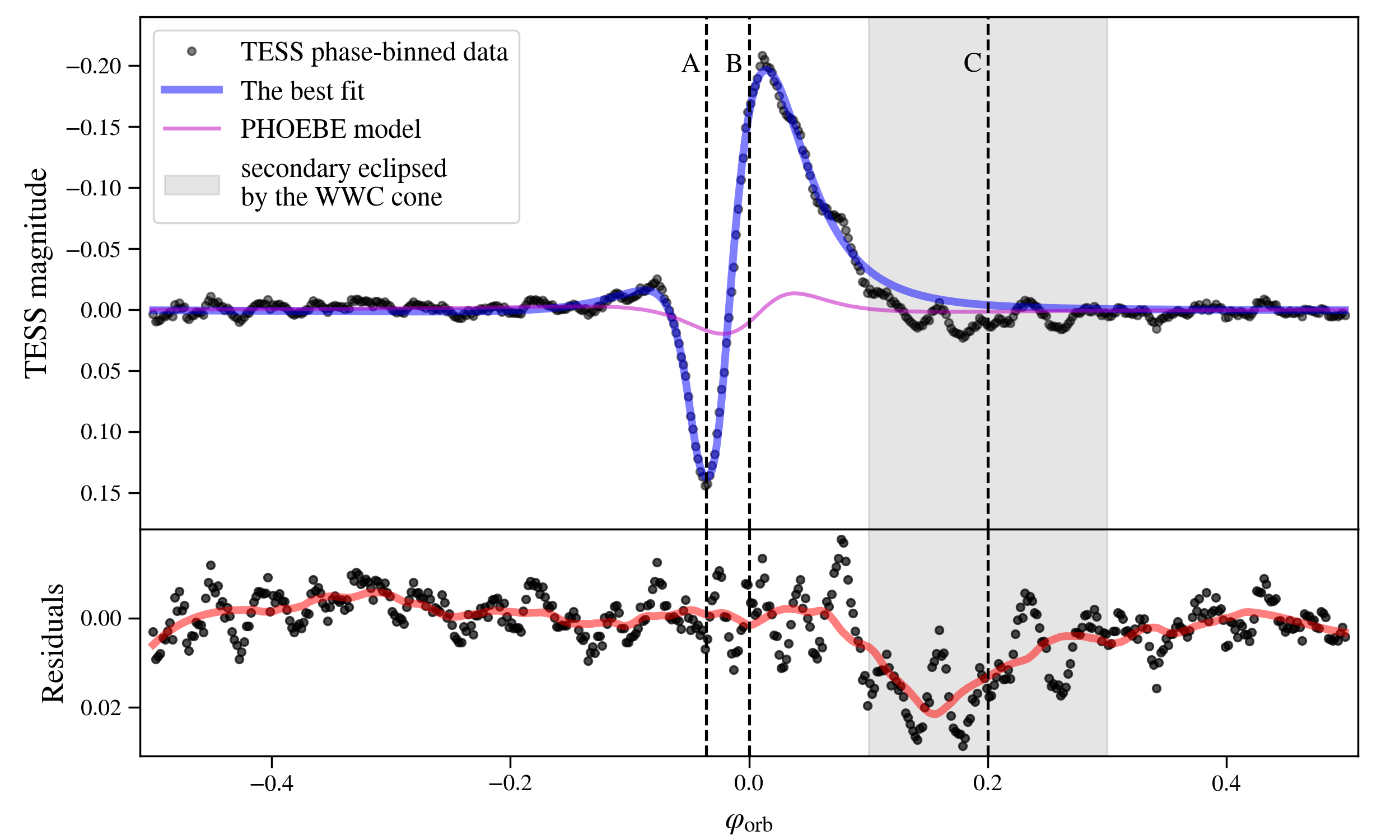}
\bigskip
\begin{minipage}{12cm}
\caption{Plot summarizing the fit of our analytical variability model to the TESS light curve of ExtEV. The upper panel shows the TESS light curve binned in the orbital phase (black dots) with the best-fit model (blue curve) superimposed. The pink line corresponds to the variability model generated with PHOEBE software \citep{2016ApJS..227...29P}. Vertical dashed lines labelled with `A', `B', and `C' mark the orbital phase of superior conjunction, periastron passage, and inferior conjunction, respectively. The vertical shaded region marks the range of orbital phases where we expect the secondary component to be partially obscured by the WWC cone. The lower panel shows the residuals from the best fit and their smoothed version (red curve). The zero phase corresponds to the time of periastron passage.}
\end{minipage}
\end{figure}

\section{Summary and conclusions}
ExtEV is not a B[e] SG because its several key properties differ significantly from those of B[e] SGs, hence its extreme brightness changes cannot be explained by a disk surrounding the primary component. It is also unlikely that the light variability in ExtEV is caused by extreme tidal distortion and subsequent nonlinear breaking of tidal waves on the surface of the primary component. The system most likely does not experience RLOF at periastron, as the rotation rate of the primary component is at most half of that assumed in the \cite{2023NatAs...7.1218M} models. Additionally, the light curve of ExtEV exhibits a similar shape and peak-to-peak amplitude in the Johnson $UBV$ and TESS passbands, ruling out extreme tidal distortions and associated gravitational darkening, which would result in heterochromatic effects in the light curve. The ellipsoidal distortion appears to play only a secondary role in shaping the light curve of ExtEV, suggesting that it is not an extreme case of a `heartbeat star'.

According to our study, ExtEV may be in fact a unique WWC binary system in which a considerable mass loss by the primary component ($4.5\times10^{-5}\,{\rm M}_\odot\,{\rm yr}^{-1}$) causes a remarkably large range of light variability. The combination of atmospheric eclipse with excess emission from the WWC cone can satisfactorily explain the amplitude and shape of the light curve, which previous models were unable to do. According to our scenario, the double-peaked emission observed in the H$\alpha$ and H$\beta$ lines should be interpreted as originating from the WWC recombination zone and the primary's stellar wind itself. The MLR of the primary component we obtained is two orders of magnitude greater than the theory predicts. We can suspect that the wind of the central BSG is tidally and rotationally enhanced, possibly with some additional interaction with TEOs. In this scenario, ExtEV is an extremely rare example of a massive binary system during a short but dramatic stage in its evolution, when the proximity of a massive companion leads to a sudden stripping of the primary's envelope without the need for RLOF. Thus we conclude that ExtEV can serve as an excellent laboratory for studying the mechanisms of wind enhancement in massive stars and predicting the impact of this enhancement on their evolution.

\begin{acknowledgments}
This research was supported by the University of Liège under the Special Funds for Research, IPD-STEMA Programme. Some of the observations reported in this paper were obtained with the Southern African Large Telescope (SALT) under program 2021-1-SCI-022 (PI: PKS). Polish participation in SALT is funded by grant No. MEiN nr 2021/WK/01. This work makes use of observations from the Las Cumbres Observatory global telescope network. PKS was supported by the Polish National Science Center grant no. 2019/35/N/ST9/03805. AP, PKS, and PŁ would like to appreciate the financial support from the Polish National Science Center grant no. 2022/45/B/ST9/03862. TR was partly founded from budgetary funds for science in 2018-2022 in a research project under the program ``Diamentowy Grant'', no. DI2018\,024648.
\end{acknowledgments}

\begin{furtherinformation}

\begin{orcids}
\orcid{0000-0003-2244-1512}{Piotr A.}{Kołaczek-Szymański}
\orcid{0009-0001-9048-7822}{Piotr}{Łojko}
\orcid{0000-0003-2488-6726}{Andrzej}{Pigulski}
\orcid{0000-0002-5819-3023}{Tomasz}{Różański}
\orcid{0000-0002-3861-9031}{Dawid}{Moździerski}
\end{orcids}

\begin{authorcontributions}
PKS was responsible for the majority of the conceptual work, data analysis, modelling, and preparation of the manuscript. PŁ developed a simplified model of emission in the H$\alpha$ and H$\beta$ lines, assuming it originates from a Keplerian disk around the primary component of ExtEV. PŁ also extracted differential photometry for ExtEV based on the data from LCOGT. AP was responsible for critically evaluating the results obtained by PKS and PŁ. He also actively participated in preparing the manuscript and responding to the reviews. TR provided normalised spectra from the SALT/HRS instrument. DM conducted an independent quality assessment of the photometry provided by PŁ.
\end{authorcontributions}

\begin{conflictsofinterest}
The authors declare no conflict of interest.
\end{conflictsofinterest}

\end{furtherinformation}

\bibliographystyle{bullsrsl-en}
\bibliography{bib}

\begin{thebibliography}{21}
\providecommand{\natexlab}[1]{#1}
\providecommand{\url}[1]{#1}
\providecommand{\urlprefix}{URL }

\bibitem[{{Bj{\"o}rklund} et~al.(2023){Bj{\"o}rklund}, {Sundqvist}, {Singh}, {Puls} and {Najarro}}]{2023A&A...676A.109B}
{Bj{\"o}rklund}, R., {Sundqvist}, J.~O., {Singh}, S.~M., {Puls}, J. and {Najarro}, F. (2023) {New predictions for radiation-driven, steady-state mass-loss and wind-momentum from hot, massive stars. III. Updated mass-loss rates for stellar evolution}.
\newblock A\&A, 676, A109.
\newblock \url{https://doi.org/10.1051/0004-6361/202141948}.

\bibitem[{{Duch{\^e}ne} and {Kraus}(2013)}]{2013ARA&A..51..269D}
{Duch{\^e}ne}, G. and {Kraus}, A. (2013) {Stellar Multiplicity}.
\newblock ARA\&A, 51(1), 269--310.
\newblock \url{https://doi.org/10.1146/annurev-astro-081710-102602}.

\bibitem[{{Fuller}(2017)}]{2017MNRAS.472.1538F}
{Fuller}, J. (2017) {Heartbeat stars, tidally excited oscillations and resonance locking}.
\newblock MNRAS, 472(2), 1538--1564.
\newblock \url{https://doi.org/10.1093/mnras/stx2135}.

\bibitem[{{Gormaz-Matamala} et~al.(2023){Gormaz-Matamala}, {Cuadra}, {Meynet} and {Cur{\'e}}}]{2023A&A...673A.109G}
{Gormaz-Matamala}, A.~C., {Cuadra}, J., {Meynet}, G. and {Cur{\'e}}, M. (2023) {Evolution of rotating massive stars with new hydrodynamic wind models}.
\newblock A\&A, 673, A109.
\newblock \url{https://doi.org/10.1051/0004-6361/202345847}.

\bibitem[{{Jayasinghe} et~al.(2021){Jayasinghe}, {Kochanek}, {Strader}, {Stanek}, {Vallely}, {Thompson}, {Hinkle}, {Shappee}, {Dupree}, {Auchettl}, {Chomiuk}, {Aydi}, {Dage}, {Hughes}, {Shishkovsky}, {Sokolovsky}, {Swihart}, {Voggel} and {Thompson}}]{2021MNRAS.506.4083J}
{Jayasinghe}, T., {Kochanek}, C.~S., {Strader}, J., {Stanek}, K.~Z., {Vallely}, P.~J., {Thompson}, T.~A., {Hinkle}, J.~T., {Shappee}, B.~J., {Dupree}, A.~K., {Auchettl}, K., {Chomiuk}, L., {Aydi}, E., {Dage}, K., {Hughes}, A., {Shishkovsky}, L., {Sokolovsky}, K.~V., {Swihart}, S., {Voggel}, K.~T. and {Thompson}, I.~B. (2021) {The loudest stellar heartbeat: characterizing the most extreme amplitude heartbeat star system}.
\newblock MNRAS, 506(3), 4083--4100.
\newblock \url{https://doi.org/10.1093/mnras/stab1920}.

\bibitem[{{Jayasinghe} et~al.(2019){Jayasinghe}, {Stanek}, {Kochanek}, {Thompson}, {Shappee} and {Fausnaugh}}]{2019MNRAS.489.4705J}
{Jayasinghe}, T., {Stanek}, K.~Z., {Kochanek}, C.~S., {Thompson}, T.~A., {Shappee}, B.~J. and {Fausnaugh}, M. (2019) {An extreme amplitude, massive heartbeat system in the LMC characterized using ASAS-SN and TESS}.
\newblock MNRAS, 489(4), 4705--4711.
\newblock \url{https://doi.org/10.1093/mnras/stz2460}.

\bibitem[{{Jermyn} et~al.(2023){Jermyn}, {Bauer}, {Schwab}, {Farmer}, {Ball}, {Bellinger}, {Dotter}, {Joyce}, {Marchant}, {Mombarg}, {Wolf}, {Sunny Wong}, {Cinquegrana}, {Farrell}, {Smolec}, {Thoul}, {Cantiello}, {Herwig}, {Toloza}, {Bildsten}, {Townsend} and {Timmes}}]{2023ApJS..265...15J}
{Jermyn}, A.~S., {Bauer}, E.~B., {Schwab}, J., {Farmer}, R., {Ball}, W.~H., {Bellinger}, E.~P., {Dotter}, A., {Joyce}, M., {Marchant}, P., {Mombarg}, J. S.~G., {Wolf}, W.~M., {Sunny Wong}, T.~L., {Cinquegrana}, G.~C., {Farrell}, E., {Smolec}, R., {Thoul}, A., {Cantiello}, M., {Herwig}, F., {Toloza}, O., {Bildsten}, L., {Townsend}, R. H.~D. and {Timmes}, F.~X. (2023) {Modules for Experiments in Stellar Astrophysics (MESA): Time-dependent Convection, Energy Conservation, Automatic Differentiation, and Infrastructure}.
\newblock ApJS, 265(1), 15.
\newblock \url{https://doi.org/10.3847/1538-4365/acae8d}.

\bibitem[{{Ko{\l}aczek-Szyma{\'n}ski} et~al.(2024){Ko{\l}aczek-Szyma{\'n}ski}, {{\L}ojko}, {Pigulski}, {R{\'o}{\.z}a{\'n}ski} and {Mo{\'z}dzierski}}]{2024A&A...686A.199K}
{Ko{\l}aczek-Szyma{\'n}ski}, P.~A., {{\L}ojko}, P., {Pigulski}, A., {R{\'o}{\.z}a{\'n}ski}, T. and {Mo{\'z}dzierski}, D. (2024) {Exploring extreme brightness variations in blue supergiant MACHO 80.7443.1718: Evidence for companion-driven enhanced mass loss}.
\newblock A\&A, 686, A199.
\newblock \url{https://doi.org/10.1051/0004-6361/202348104}.

\bibitem[{{Ko{\l}aczek-Szyma{\'n}ski} et~al.(2022){Ko{\l}aczek-Szyma{\'n}ski}, {Pigulski}, {Wrona}, {Ratajczak} and {Udalski}}]{2022A&A...659A..47K}
{Ko{\l}aczek-Szyma{\'n}ski}, P.~A., {Pigulski}, A., {Wrona}, M., {Ratajczak}, M. and {Udalski}, A. (2022) {Tidally excited oscillations in MACHO 80.7443.1718: Changing amplitudes and frequencies, high-frequency tidally excited mode, and a decrease in the orbital period}.
\newblock A\&A, 659, A47.
\newblock \url{https://doi.org/10.1051/0004-6361/202142171}.

\bibitem[{{Lamers} et~al.(1998){Lamers}, {Zickgraf}, {de Winter}, {Houziaux} and {Zorec}}]{1998A&A...340..117L}
{Lamers}, H. J.~G.~L.~M., {Zickgraf}, F.-J., {de Winter}, D., {Houziaux}, L. and {Zorec}, J. (1998) {An improved classification of B[e]-type stars}.
\newblock A\&A, 340, 117--128.

\bibitem[{{Lamontagne} et~al.(1996){Lamontagne}, {Moffat}, {Drissen}, {Robert} and {Matthews}}]{1996AJ....112.2227L}
{Lamontagne}, R., {Moffat}, A. F.~J., {Drissen}, L., {Robert}, C. and {Matthews}, J.~M. (1996) {Photometric Determination of Orbital Inclinations and Mass Loss Rates for Wolf-Rayet Stars in WR+O Binaries}.
\newblock AJ, 112, 2227.
\newblock \url{https://doi.org/10.1086/118175}.

\bibitem[{{MacLeod} and {Loeb}(2020)}]{2020ApJ...902...85M}
{MacLeod}, M. and {Loeb}, A. (2020) {Hydrodynamic Winds from Twin-star Binaries}.
\newblock ApJ, 902(1), 85.
\newblock \url{https://doi.org/10.3847/1538-4357/abb313}.

\bibitem[{{MacLeod} and {Loeb}(2023)}]{2023NatAs...7.1218M}
{MacLeod}, M. and {Loeb}, A. (2023) {Breaking waves on the surface of the heartbeat star MACHO 80.7443.1718.}
\newblock Nature Astronomy, 7, 1218--1227.
\newblock \url{https://doi.org/10.1038/s41550-023-02036-3}.

\bibitem[{{MacLeod} et~al.(2022){MacLeod}, {Vick} and {Loeb}}]{2022ApJ...937...37M}
{MacLeod}, M., {Vick}, M. and {Loeb}, A. (2022) {Tidal Wave Breaking in the Eccentric Lead-in to Mass Transfer and Common Envelope Phases}.
\newblock ApJ, 937(1), 37.
\newblock \url{https://doi.org/10.3847/1538-4357/ac8aff}.

\bibitem[{{Paxton} et~al.(2011){Paxton}, {Bildsten}, {Dotter}, {Herwig}, {Lesaffre} and {Timmes}}]{2011ApJS..192....3P}
{Paxton}, B., {Bildsten}, L., {Dotter}, A., {Herwig}, F., {Lesaffre}, P. and {Timmes}, F. (2011) {Modules for Experiments in Stellar Astrophysics (MESA)}.
\newblock ApJS, 192(1), 3.
\newblock \url{https://doi.org/10.1088/0067-0049/192/1/3}.

\bibitem[{{Pr{\v{s}}a} et~al.(2016){Pr{\v{s}}a}, {Conroy}, {Horvat}, {Pablo}, {Kochoska}, {Bloemen}, {Giammarco}, {Hambleton} and {Degroote}}]{2016ApJS..227...29P}
{Pr{\v{s}}a}, A., {Conroy}, K.~E., {Horvat}, M., {Pablo}, H., {Kochoska}, A., {Bloemen}, S., {Giammarco}, J., {Hambleton}, K.~M. and {Degroote}, P. (2016) {Physics Of Eclipsing Binaries. II. Toward the Increased Model Fidelity}.
\newblock ApJS, 227(2), 29.
\newblock \url{https://doi.org/10.3847/1538-4365/227/2/29}.

\bibitem[{{Sana} et~al.(2012){Sana}, {de Mink}, {de Koter}, {Langer}, {Evans}, {Gieles}, {Gosset}, {Izzard}, {Le Bouquin} and {Schneider}}]{2012Sci...337..444S}
{Sana}, H., {de Mink}, S.~E., {de Koter}, A., {Langer}, N., {Evans}, C.~J., {Gieles}, M., {Gosset}, E., {Izzard}, R.~G., {Le Bouquin}, J.~B. and {Schneider}, F.~R.~N. (2012) {Binary Interaction Dominates the Evolution of Massive Stars}.
\newblock Science, 337(6093), 444.
\newblock \url{https://doi.org/10.1126/science.1223344}.

\bibitem[{{Shenar} et~al.(2021){Shenar}, {Sana}, {Marchant}, {Pablo}, {Richardson}, {Moffat}, {Van Reeth}, {Barb{\'a}}, {Bowman}, {Broos}, {Crowther}, {Clark}, {de Koter}, {de Mink}, {Dsilva}, {Gr{\"a}fener}, {Howarth}, {Langer}, {Mahy}, {Ma{\'\i}z Apell{\'a}niz}, {Pollock}, {Schneider}, {Townsley} and {Vink}}]{2021A&A...650A.147S}
{Shenar}, T., {Sana}, H., {Marchant}, P., {Pablo}, B., {Richardson}, N., {Moffat}, A.~F.~J., {Van Reeth}, T., {Barb{\'a}}, R.~H., {Bowman}, D.~M., {Broos}, P., {Crowther}, P.~A., {Clark}, J.~S., {de Koter}, A., {de Mink}, S.~E., {Dsilva}, K., {Gr{\"a}fener}, G., {Howarth}, I.~D., {Langer}, N., {Mahy}, L., {Ma{\'\i}z Apell{\'a}niz}, J., {Pollock}, A.~M.~T., {Schneider}, F.~R.~N., {Townsley}, L. and {Vink}, J.~S. (2021) {The Tarantula Massive Binary Monitoring. V. R 144: a wind-eclipsing binary with a total mass >140 Mo}.
\newblock A\&A, 650, A147.
\newblock \url{https://doi.org/10.1051/0004-6361/202140693}.

\bibitem[{{Smith}(2014)}]{2014ARA&A..52..487S}
{Smith}, N. (2014) {Mass Loss: Its Effect on the Evolution and Fate of High-Mass Stars}.
\newblock ARA\&A, 52, 487--528.
\newblock \url{https://doi.org/10.1146/annurev-astro-081913-040025}.

\bibitem[{{Vanbeveren} et~al.(1998){Vanbeveren}, {De Donder}, {Van Bever}, {Van Rensbergen} and {De Loore}}]{1998NewA....3..443V}
{Vanbeveren}, D., {De Donder}, E., {Van Bever}, J., {Van Rensbergen}, W. and {De Loore}, C. (1998) {The WR and O-type star population predicted by massive star evolutionary synthesis}.
\newblock New Astronomy, 3(7), 443--492.
\newblock \url{https://doi.org/10.1016/S1384-1076(98)00020-7}.

\bibitem[{{Vink}(2022)}]{2022ARA&A..60..203V}
{Vink}, J.~S. (2022) {Theory and Diagnostics of Hot Star Mass Loss}.
\newblock ARA\&A, 60, 203--246.
\newblock \url{https://doi.org/10.1146/annurev-astro-052920-094949}.

\end{thebibliography}

\end{document}